# Using an Educational IoT Lab Kit and Gamification for Energy Awareness in European Schools

**Georgios Mylonas, Dimitrios Amaxilatis, Lidia Pocero, Iraklis Markelis**
Computer Technology Institute & Press "Diophantus"
Patras, Greece
mylonasg@cti.gr

**Joerg Hofstaetter**
OVOS
Vienna, Austria
jho@ovos.at

**Pavlos Koulouris**
Ellinogermaniki Agogi,
Athens, Greece
pkoulouris@ea.gr

**ABSTRACT**
The use of maker community tools and IoT technologies inside classrooms is spreading in an increasing number of education and science fields. GAIA is a European research project focused on achieving behavior change for sustainability and energy awareness in schools. In this work, we report on how a large IoT deployment in a number of educational buildings and real-world data from this infrastructure, are utilized to support a "maker" lab kit activity inside the classroom, together with a serious game. We also provide some insights to the integration of these activities in the school curriculum, along with a discussion on our feedback so far from a series of workshop activities in a number of schools. Our initial results show strong acceptance by the school community.

## INTRODUCTION

Raising awareness among young people and changing their habits concerning energy usage is considered key in achieving a sustainable energy behavior. Education ministries manage thousands of buildings, while also educational institutions revolve around students, educators, and policy makers, all of which are essential in implementing energy-awareness policies. Furthermore, EU considers environmental education one of the most prominent instruments to influence human behavior towards sustainability [1], while educational buildings constitute 17% of the EU non-residential building stock (in $m^2$) [2]. In top of these, evidence shows that a focus on energy use in schools results in multiple benefits, along with educational excellence and a healthy learning environment [3].

Moreover, a key objective of energy efficiency initiatives in schools is making students aware that energy consumption is influenced by the sum of individual behaviors and that simple behavior changes and interventions have a tangible impact towards energy savings. To this end, IoT technologies can support such initiatives with immediate feedback regarding the impact of our actions and automating the implementation of energy-savings policies, while maintaining comfort levels. The availability of actual measurements of environmental parameters, such as energy consumption, indoor and outdoor luminosity, temperature, noise, and so on, enables the conception and realization of a number of diverse education-related applications and scenarios: teachers can use collected data and analytics during the class to explain basic phenomena related to the parameters monitored, or they organize projects where students monitor such parameters at their class or home.

Green Awareness In Action – GAIA [4], a Horizon2020 EC-funded project, is developing an IoT platform that combines sensing, web-based tools and gamification elements, in order to address the educational community. Its aim is to increase awareness about energy consumption and sustainability, based on real-world sensor data produced by the school buildings where students and teachers live and work, while also lead towards behavior change in terms of energy efficiency.

In this context, we believe an approach based on open-source, replicable and widely available technology, providing a "foundation" that lets educators to adapt to the needs of each class, opens up many possibilities. We base this approach on the assumption that, by using this hands-on approach, it will allow for a more personalized and engaging experience to the students.

We discuss here an educational lab kit that utilizes the IoT infrastructure of GAIA as an input generator, i.e., real-world data, which has been developed over open-source technologies. The approach followed means it is replicable for schools wishing to combine science/technology classes with a twist on energy efficiency and sustainability. We present our design for this lab kit, its integration with a gamification component and some positive preliminary end-user evaluation results.

We continue outlining recent related work, followed by a quick introduction to GAIA and its overall philosophy. The educational lab kit of GAIA is then presented in more detail, along with a discussion on GAIA Challenge, the gamification component of GAIA, in the following section. We also present some preliminary results from a series of workshops with primary and secondary school students in Greece, before concluding this work in the las section.

**RELATED WORK**

A significant number of research projects and activities focus on the energy efficiency domain. BuildUp [5] is a European portal presenting in a systematic manner many related approaches. [6] and [7] are examples of projects that involve the educational community and energy efficiency, but which also relate to building retrofitting with new materials, etc. Our work sees energy awareness through the prism of behavioral change towards more sustainable practices in school buildings.

There are also a number of past projects, like SEACS [8], that produced material related to sustainability and STEM, or on-going ones like UMI-Sci-Ed [9], that tie educational content to the Internet of Things. However, very few attempts are made to tie together real-world IoT data with a curriculum aiming for long-term behavioral change, like in the case of GAIA.

Furthermore, a number of recent projects like Charged [10], Entropy [11], GreenSoul [12], Tribe [13], target diverse end-user communities and do not focus on the educational community. Although to a certain extent their target users belong to schools or universities, the apps, serious games and content they have developed does not aim to provide an educational tool tied to a school's curriculum. In our work, we try to provide a kit that will help towards energy awareness, but also appear attractive to educators as a tool to make their classes more interactive and engaging.

**THE GAIA PROJECT**

GAIA utilizes a number of IoT installations in school buildings in Greece, Italy and Sweden. To change the behavior and achieve sustainable results, GAIA utilizes a loop-based approach focused around 3 pillars: raise awareness, support action, and foster engagement. Each installation consists of a multitude of IoT nodes, which communicate with a cloud infrastructure via a gateway device. Such nodes comprise multiple sensing devices, while the gateway nodes coordinate communication and enable interaction with cloud-based services [14].

The overall design pattern for the installation of IoT infrastructure in the school buildings participating in the project, is as follows:

- Nodes installed to monitor the power consumption of the building as a whole, or specific floors/sectors.
- Classroom nodes monitoring parameters such as temperature, humidity, motion and noise levels.
- Gateway nodes bridging IoT nodes that use IEEE 802.15.4 with the Internet. The IoT gateways communicate directly with GAIA cloud services.

The hardware design of the IoT nodes utilized in GAIA follows an open-source approach [15], using hardware components widely available, paired with some custom designs, for example custom PCBs for interconnecting sensors to MCUs, etc. Almost all available infrastructure is based on Arduino-compatible or Raspberry Pi components. GAIA releases the specifications for this infrastructure and related results as open-source, in all cases wherever possible. For this purpose, a series of related material is available at the project's GitHub repository[1].

Based on this IoT infrastructure, we monitor energy consumption and environmental parameters inside school buildings that participate in the project. Such data are used as input to the educational aspects of the project. The idea is to provide students and educators with a more "personalized" approach: GAIA's educational material and applications will use the actual situation in the school buildings to make interaction more engaging, fun and useful. Since the hardware used is open-source and components are commercially available, schools and students have the option to purchase the components and *assemble* the IoT nodes themselves. In fact, one of the schools participating in the project has already done exactly this. The school, a technical high school/college in Sweden, purchased the required hardware, and its students assembled and installed the devices in their building, under the supervision of the school staff.

---

[1] GAIA project repository, https://github.com/GAIA-project

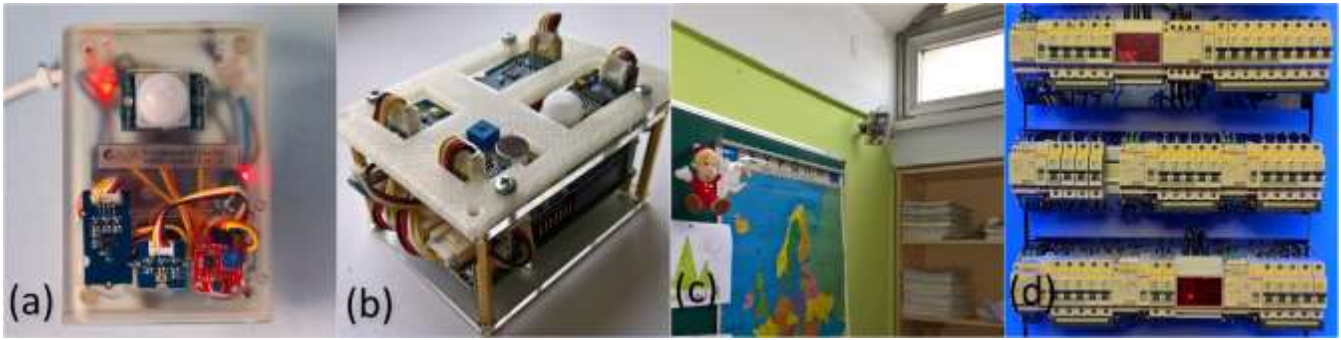

**Figure 1. Sample photos of the device lineup inside GAIA's schools: a) an Arduino-based IoT node for classrooms, b) a Raspberry Pi-based node for classrooms, c) an IoT node inside a classroom, d) power meters installed on electricity distribution panels.**

**THE GAIA EDUCATIONAL LAB KIT**

In short, the kit aims to teach students using a "hands-on" approach, in which they get to use IoT components and electronics. Based on guides provided by the project, they examine data from their school building and go through the peculiarities of consuming energy, how the building behaves in the various classrooms in terms of environmental parameters, and more. The kit includes already assembled devices and commercial IoT sensors and actuators to allow students complete classes and lab tutorials regarding energy and sustainability, as well as provides guidelines for implementing crowdsensing quests (also related to the gamification component of GAIA). In such quests, students create a "map" of specific parameters, e.g., energy, insulation, in their school building. It also serves as an additional means of interacting with the project and further increasing the end-user engagement, along with the other tools of GAIA, such as the gamification platform and the BMA.

In the majority of the available educational lab kit activities, the focus is on mapping what is actually happening inside the school building in real time. This is achieved in two ways:

- Using a floorplan of their school as a surface to place electronic components and devices, essentially assembling a small-scale interactive installation.
- Conducting simple "missions" inside the different parts of their buildings, indoors and outdoors, detecting and pinpointing specific energy-efficiency-related rooms or devices.

The Lab Kit includes IoT devices, commercial or GAIA-designed, together with other hardware components that can be used to assemble custom electrical circuits, which can enable:

- *Familiarization with electronics*, using common components, such as LEDs, resistors, switches, etc.
- *Monitoring of real-world parameters*, in a way similar to the fixed IoT infrastructure installed inside GAIA school buildings, but in a more "personalized" and direct manner.
- *Interaction with GAIA software services*, in similar ways to the ones used by the infrastructure, also providing a clearer picture of how the overall GAIA system works.

We have avoided the use of aspects such as extensive wiring between the different components, or complicated interconnections by using:

- *Conductive ink*, in order for students to "draw" the interconnections required between the various electronic components. In this manner, we avoid the extensive use of cables and breadboards, simplifying the overall activity and saving time.
- *Raspberry GrovePi hats,* that have standardized connectors for input/output interfaces, i.e., students and teachers do not have to worry about how to connect the components, or where, since all interconnections are carefully labeled to prevent confusion.
- *Magnet-carrying electronic components*, which enable easy placement on top of a metal surface. The use of magnets prevents the components from slipping and moving around during the lab kit activity.
- *Printed paper school building floorplans,* which illustrate both the parts of the building and the placement of electronic components for the lab activity. The floorplans are placed on top of a metal surface so that magnets are kept firmly connected during the lab activity.

Regarding the actual bill of materials of the Lab Kit, we utilize the following components:

- *Raspberry Pi*, as the device handling the main computing and networking duties for the lab activities.

- *Conductive ink markers*, for sketching out wire paths onto paper. At this point, we are using the Circuitscribe[2] markers.

- *Electronic components*, such as LCD screens, resistors, switches, potentiometers, etc. At this point, we are using Circuitscribe components, mainly due to their magnetic clip attachments, ease of use and overall safety.

- *Sensors with standardized interfaces for connecting to the Raspberry Pi*. At this point, we are using the GrovePi[3], an *open-source family of sensors* available from various distributors, which use standardized interfaces and are suitable for educational activities due to their design.

- *Custom electronic boards*, which ease interfacing with GAIA and visualization of real-time data used during the lab activities. One of such boards is the LED ring board used to visualize the 3-phase energy consumption in the building (Figure 7).

- We plan to enhance the available options and use additional components and technologies in the coming months, to provide more options for interfacing as well as cover cases where schools may already have similar equipment and would like to utilize it.

Regarding the software part of the Lab Kit, the Raspberry Pi devices use the Raspbian Linux distribution, while the Lab Kit activities are based on a number of Python scripts provided by the consortium. The lab kit activities use the default options for Python provided by Raspbian, while we use the standardized Web interfaces of GAIA for communicating to the system for real-time data, thus no additional changes are required to the vanilla Raspbian distribution for Raspberry Pi.

**Organization of the Lab Kit activities**

Regarding the overall organization of the activities and the provided material, the consortium has prepared a series of lab activities, covering aspects of energy consumption and efficiency inside school buildings. The thematic list covered is the following:

- "Energy consumption in our school".
- "Lighting inside school buildings".
- "Heating inside school buildings".
- "Temperature, Humidity and Thermal Comfort".
- "Devices and Energy efficiency".
- "Energy Inspectors - The energy footprint of our building".

An additional activity can be implemented in case a certain school would like to implement a "stable" interactive installation, in the form of a class project by the students, in order to depict some kind of energy efficiency metric in its own school building. We include some examples of activities in the annex of this document.

Regarding the provided material, there are available guides for each activity. In the description of each activity, we include the title of the subject, the necessary cognitive background for the teams (theoretical and practical) and a short description of the tasks to be completed (goal). One set of material concerns the educators, identifying the educational target for each activity, the methods used, as well as a schedule for the proposed lab activity. Another set of material addresses the students' part, giving specific instructions on how to perform the envisioned activities, explaining how to interconnect sensors and electronic components, and how to execute the Python scripts provided by the project.

**Summary of a sample lab kit activity**

We present here briefly the activity "Temperature, Humidity and Thermal Comfort". Students are given a short introduction to the aims of the activity, as well as instructions on how to draw circuits for the lab using conductive ink, on top of a printed floorplan of their school. They are then instructed to assemble a small electronic circuit using 2-color LEDs, buttons and an LCD screen, and place the components over a predefined set of class rooms in the floor map. After assembling the circuit, students power up the Raspberry Pi's, and start looking into the Python code. They execute a series of available Python scripts that connect to GAIA's cloud infrastructure to fetch real-time data. They then go through a series of activities, where they see the temperature and humidity inside their classrooms, which are visualized on the LEDs of the circuit (e.g., red for temperatures above 25 degrees) and the LCD screen.

They use hardware switches and buttons to navigate between different visualization modes. A set of activities on the instructions provides the background to assess thermal comfort levels, to note down differences between classroom, and try to reason the origin of such differences due to room orientation, construction, etc. They also provide examples of how to customize the Python code to provide additional functionality and visualization modes.

---

[2] https://www.circuitscribe.com/

[3] https://github.com/DexterInd/GrovePi

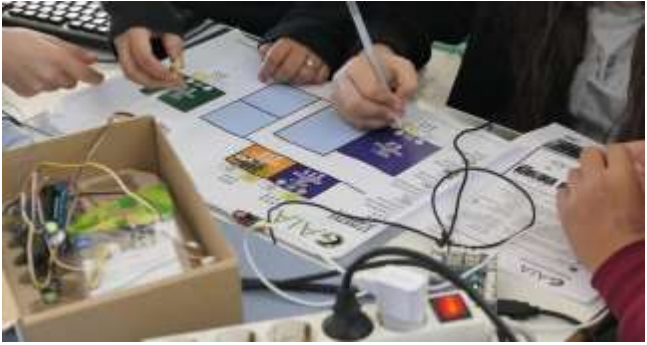

**Figure 2. Example of an in-class activity with the Lab Kit. Primary school students use conductive ink to draw circuits on top of a printed floor map of their school.**

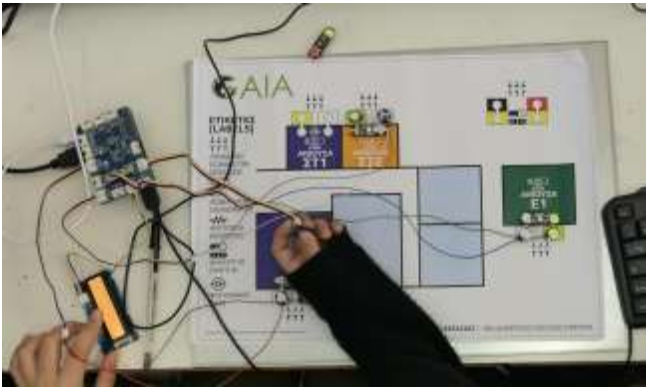

**Figure 3. Example of an in-class activity with the Lab Kit. A student uses the kit to visualize temperature readings in different classrooms of her school building (purple, orange and green rooms)**

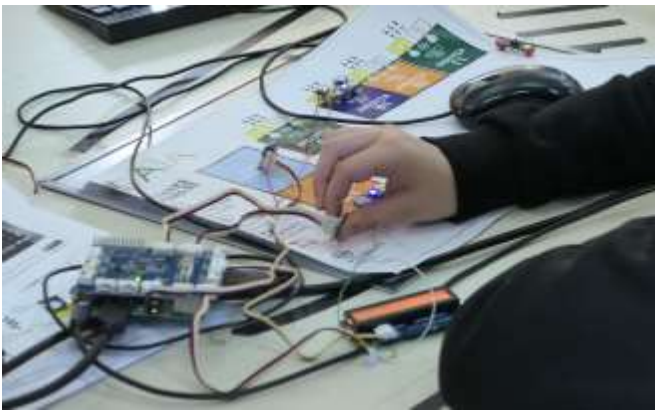

**Figure 4. Example of an in-class activity with the Lab kit. A student uses a button to visualize different modes of readings inside his school on LEDs and an LCD screen.**

## THE GAIA CHALLENGE

GAIA Challenge is an online serious game which raises students' and teachers' energy awareness within their own facility, accessible through web browsers. The challenge utilizes gamification mechanics to:

- motivate participants to engage in energy saving topics,
- work on online "quests",
- participate in real-life activities,
- experience their impact on the facilities' energy consumption over the course of the challenge,
- compete and compare against other classes and schools in other countries,
- share their experiences with their peer group.

Real-time data from sensors in the buildings are used as part of the challenge, in order to visualize the real life impact of the participants' behavior and build collaborative (within a facility) and competitive (between facilities) gamification elements. Teachers are invited in Class Activities to work together with their students on hands-on observation and optimization tasks in classrooms. The online challenge offers the following core features:

*Quest Map:* The Quest Map is the main view for the user. It is an interactive visual representation of multiple entities of the online challenge. It symbolizes the user's journey from the start (top) to the finish line (bottom) of the challenge. Each Quest Node is related to a specific topic. Along the way there are also Quest Sequences and Class Activities, with multiple quests for the user to play. These quests are grouped into five subject areas related to energy consumption reduction. There are also bonus areas with quests available for classes that participate in Class Activities.

*Class Activities:* Class Activities are crossovers of learning in class, engaging in the online challenge and on-site engagement in the facility, requiring a teacher and on-site engagement. The teacher can start a Class Activity for her class at any time during the online challenge. A Class Activity is divided into three parts: (a) learning the theory, (b) consolidating the knowledge and (c) applying it. Teachers can decide on the topic, the actual activities and the physical space of these Class Activities.

*Community Dashboard:* users can compare the performance and score between classes and schools. The user can select a class from the list and inspect its achievements, its user ranking and also a list of recent snapshots by the class' students is shown. The class score is the sum of points of the class' students.

### Integration with the Lab Kit

The lab kit activities have been linked with GAIA Challenge, in the form of additional quests unlocked for the schools that are planning or are participating in such activities. Essentially, these quests are a visual "tutorial" for students to prepare for the lab or understand better aspects such as e.g., the wiring between specific electronic components. Educators can choose to "unlock" these Lab Kit-specific quests before or after the first Lab Kit activity, in order to familiarize students with the Lab, or repeat some of the activities to get a better understanding of the tools used.

We also provide sample additional activities to the educators, suggesting ways to integrate the rest of the GAIA toolkit with the lab activities, e.g., in the period between two successive lab activities. students can use the Building Manager Application to monitor environmental parameters that they checked during a lab activity. They can then discuss their findings with their teachers when they have their next lab activity at school. They are also encouraged to share their findings/progress on the GAIA Challenge or Facebook page of the project. GAIA Challenge can also be used as homework, so that students do not lose contact with the workshops for a long time.

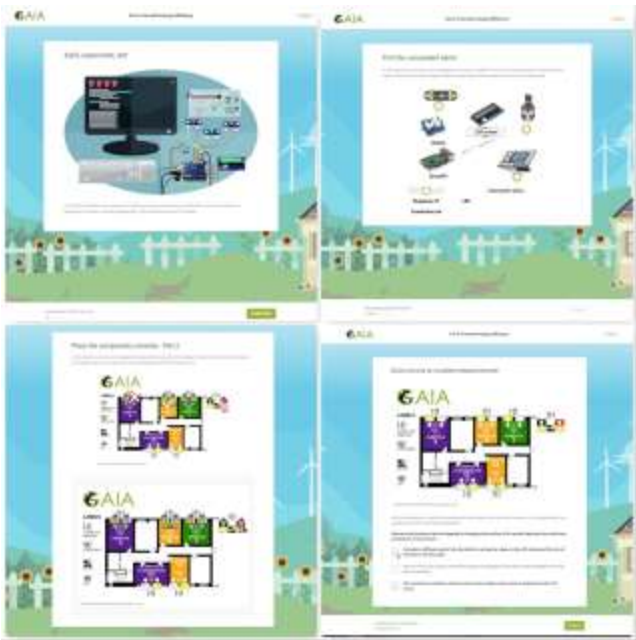

**Figure 5. Screenshots for the Lab Kit mission included in the GAIA Challenge.**

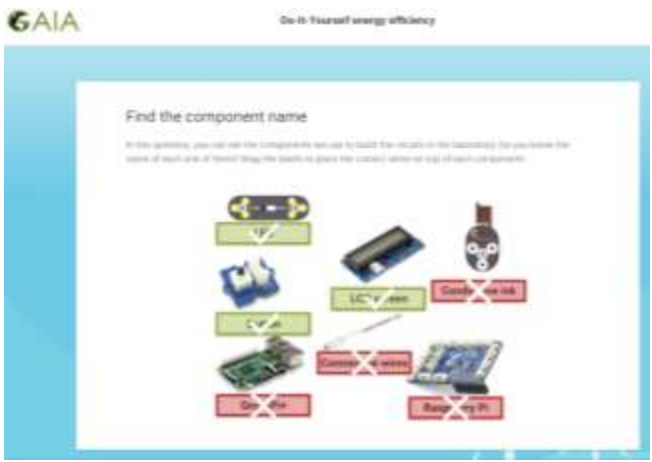

**Figure 6. Example of activity within GAIA Challenge, asking students to identify the names of the components used with the Lab Kit.**

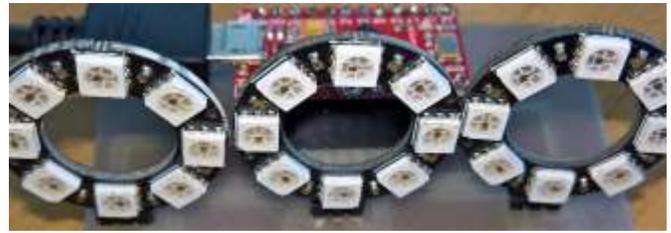

**Figure 7. Example of a custom component used in some of the Lab Kit activities. The 3 LED rings are used as "dials" to display levels of power consumption, temperature, humidity.**

**INITIAL END-USER EVALUATION AND DISCUSSION**

In this section, we report on some preliminary results we have from the use of the lab kit in schools in Greece the past months. Since the lab kit is still an ongoing work, we report our experiences from its initial application inside the curriculum of a number of schools. It is important to note that several of the schools participating in the project have educational staff that utilizes, or is comfortable with the use of e.g., Arduino-based activities. However, not all educators or students have experience with such technologies. Thus, it is important to stress that the discussion here applies to a diverse set of end-users.

For a preliminary evaluation of the lab kit inside classrooms, we conducted 2 workshops with 2 groups of Greek students aged 11-15 years old (primary and secondary school students), including 48 and 58 students in each workshop.

The questions asked were the following:

- Did you have prior experience with electronic circuits before the lab (*Y/N*)?
- Did you face serious difficulty completing the activities during the lab (*Y/N*)?
- Do you think such an activity will help you learn something about your school building and energy consumption (grade 1-5)?
- If possible, would you consider repeating similar activities at your home (grade 1-5)?
- Did you enjoy the lab activity overall? (grade 1-5)?

With respect to the answers given by the students, there was a very positive response from the students. In the first lab, 71% answered that they liked very much the activity (5/5), 23% gave out 4/5, and the rest 3/5. In this case, 58% had some sort of previous contact with electronics, while 87.5% stated that they did not face difficulties during the lab. 98% said that they thought such an activity could help them to learn something about energy in buildings, while 94% said that they would consider doing such an activity at home, if possible. In the second lab, there was some change to the positive response, with 58% rating it 5/5, 26% 4/5, and 12 3/5, possibly also reflecting the change of topic for the lab. An 80% stated that they did not face difficulties during the lab, while 96% thought that they learned useful about their building.

Another group consisting of 7 primary and secondary school teachers in Greece, participated in a daily workshop showcasing some sample lab kit activities. After completing the workshop, an evaluation questionnaire was given to them. The questions asked in this group were:

- Did you find the activity and instructions clear (*Y/N*)?
- Do you believe you could conduct the activities at your school without help from GAIA (*Y/N*)?
- Did you face difficulties in the activities (*Y/N*)?
- Do you believe that students will gain/learn something out of this activity (*Y/N*)?

Regarding their answers, 5 teachers answered that they found the activity and instructions 'quite clear', 1 'absolutely clear' and 1 'clear enough'. Regarding difficulties faced, only 1 teacher commented that enough time should be given to complete the activities, depending on the conditions in each school. All teachers answered that they thought that students could learn something by engaging in such activities.

Thus, so far, we have had a very positive overall response from both students and teachers. The initial results reported in this section will be further strengthened in the following weeks and months, as lab kit activities are carried out in the vast majority of the schools participating in GAIA. We expect close to 15 schools to have a first experience with the lab kit in the near future, and several hundreds of students to be involved in them.

**CONCLUSION**

We presented here an educational lab kit that uses real-time data from school buildings, in order to enable a more personalized approach to educational activities for energy awareness and sustainability. This is included as part of the educational approach of GAIA, which uses real-world data from school buildings to provide a more hands-on and personalized experience aspect in class activities. As we are progressing in the coming months with our implementation of the lab activities within classrooms, we will also be looking into actual results with respect to behavioral change and energy efficiency, in combination with the main body of educational activities in the project. It is already clear that activities such as the lab kit have potential as tools for teaching sustainability concepts and science topics.

**ACKNOWLEDGMENTS**

This work has been partially supported by the EU research project "Green Awareness In Action", funded by the EC and EASME under H2020 and contract number 696029.